\documentstyle[12pt]{article}
\input epsf
\begin{document}
\begin{titlepage}

\vspace{40mm}

\centerline{\large \bf Hadronic component of the photon}
\vspace{5mm}
\centerline{\large \bf spin dependent structure function $g_1^{\gamma}$
from QCD}
\vspace{10mm}

\centerline{\bf A V Belitsky}
\vspace{10mm}

\centerline{\it Bogoliubov Laboratory of Theoretical Physics}
\vspace{3mm}
\centerline{\it Joint Institute for Nuclear Research}
\vspace{3mm}
\centerline{\it 141980, Dubna, Russia}
\vspace{20mm}

\vspace{2cm}

\centerline{\bf Abstract}

We calculate the spin dependent structure function of the polarized
virtual photon $g_1^{\gamma}(x,Q^2,p^2)$, especially its hadronic
part, using the OPE in the inverse powers of the target photon
virtuality. Some model is accepted to achieve a correct analytical
behaviour in the photon squared mass. This enables extrapolation to
the case of a real photon.

\end{titlepage}
\newpage

\section{Introduction}

More than two decades ago it was observed \cite{ter73}
that the processes $e^{\pm}e^-\to e^{\pm}e^-X$ being summed
over all hadronic states (see fig. 1.) allowed for the study
of $\gamma\gamma$ interactions at high energies. An experimentally
favourable situation for mea\-su\-re\-ments of this cross section
appears when one electron (corresponding to the photon with
momentum $q$) is detected at large momentum transferred $Q^2=-q^2$
and the other remains in a nearly forward domain
$-p^2\approx 0$. So, in practice, it is more appropriate
to deal with the deep inelastic scattering (DIS) of an electron
beam on a real photon target which is characterized by several
nonpolarized and polarized structure functions in the same way
as the usual DIS on a nucleon.

The photon structure functions in the context of the gauge theories
have first been discussed by Ahmed and Ross \cite{ahm75}. Since
1977 \cite{wit77} the deep inelastic electron scattering from
a photon target has become a new subject of intensive theoretical
studies in the framework of QCD. Much progress has been achieved
in this direction \cite{vdm}. As was firstly advocated by Witten
the nonpolarized structure function $F_2(x)$ can be calculated
using the perturbative QCD alone. At that time there was considerable
optimism that this process was an excellent test for the perturbative
QCD and might provide an accurate measurement of ${\alpha}_s$.
By now the optimism has waned considerably. This happens because
Witten's suggestion is true only for the asymptotically large
probe-photon momentum transfer squared where a "contact"-type term
due to the photon operators in the framework of the Operator Product
Expansion (OPE) turns out to
be dominant. At smaller values of $Q^2$ the hadronic component
becomes sizable, the photon admits considerable contribution that
comes from the nonperturbative region. The photon has a partonic
substructure induced by virtual fluctuation into the quark-antiquark pair.
This process receives the short as well as long distance contribution,
and the latter is due to the nonperturbative dynamics that cannot be
calculated in perturbative QCD. Only if $-p^2$ is in the deep
inelastic scattering region, the partonic substructure can be neglected
and the photon exhibits its point-like nature. For finite $-p^2$ the
target photon receives contribution from the hadronic component associated
with the Vector Meson Do\-mi\-nance (VMD) of the soft photon. Due to this
contamination the hope of pure extraction of ${\Lambda}_{QCD}$
from this reaction in a real experiment fails owing to the large
uncertainty in the theoretical prediction for this part. Till
recently the only es\-ti\-ma\-tions for the latter have been obtained
from the VMD models \cite{vdm}.

The first QCD based calculation of the hadronic part has been initiated
by Balitsky \cite{bal82}. However, only a few first moments rather than
$x$-dependence of the structure function were found in his paper.
Recently a new approach for calculating the photon
structure function in QCD has been developed \cite{iof90}. It enables one
to evaluate the structure function in the region of intermediate $x$
and was successfully applied to the case of spin averaged scattering.

In recent times the polarized photon structure functions
have attracted a lot of attention. In ref. \cite{man89} OPE and
the renomalization group analysis were extended to the polarized
sector, while works cited in ref. \cite{ter89} deal with the first
moment of the spin dependent structure function $g_1^{\gamma}(x)$
and its sensitivity to the chiral symmetry realization. In these
papers, as in Witten's one, the hadronic component was completely
disregarded. However, the knowledge of the latter is very important
for comparison of the theoretical predictions with a real experiment.
This holds for all presently available or foreseeable values of $Q^2$.

In the present study we concentrate on the calculation of
the polarized structure function $g_1^{\gamma}(x)$ following
the method mentioned above. We shall start with the consideration
of the structure function when the target photon virtuality is
large and spacelike $-p^2\gg R_c^{-2}$, $R_c$ is a confinement radius,
but $-p^2\ll Q^2$ and apply the OPE to the discontinuity \cite{iof85}
of the forward photon-photon scattering amplitude which results in the
expansion in the inverse powers of target photon virtuality.
On the other hand, using the analytical properties of the structure
function in the photon off-shellness $-p^2$ it is represented
via the dispersion relation through the contribution of the vector meson
and continuum. Comparing the two representations of the same quantity
we can fix unambiguously all parameters of hadronic spectrum.
The structure function obtained possesses correct analytical
properties in the target photon mass and has no fictitious kinematical
singularities inherent in the per\-tur\-ba\-tive diagrams.

The sensitivity to the QCD radiative corrections is poor until
very large $Q^2$ is attained. Thus, we restrict ourselves to the
lowest order graphs and to the consideration of light quarks only;
the result is correct for intermediate $Q^2\leq 10GeV^2$. However,
for too low $Q^2$ the higher twist corrections may be important.

\section{OPE for the four-point amplitude}

To start with we consider the four-point correlation function
\begin{equation}
T_{\mu \nu , \alpha \beta}=4\pi {\alpha}_{em}i^3
\int d^4xd^4yd^4z e^{iqx+ip(y-z)}
\langle 0|T\{ j_{\mu}(x)j_{\nu}(0)j_{\alpha}(y)j_{\beta}(z)\}|0\rangle,
\end{equation}
where $j_{\mu}=\sum_{q}Q_q \bar{\psi}_q {\gamma}_{\mu} {\psi}_q$ is
the electromagnetic quark current. This amplitude is originated from
the $T$-product of two electromagnetic currents between the photon
states and application of the Lehmann-Symanzik-Zimmermann reduction
formula. The discontinuity across the branch cut on the real axis in
the complex plane of $\omega =\frac{1}{x}$, where $x$ is the usual
Bjorken variable, gives us the structure function we are interested in.

Note that due to the weak coupling of the photon to the hadronic
states there is no need to extract the physical state of interest
with the help of some auxiliary procedure, {\it e.g.} the Borel
transformation.

In order to find the polarized spin structure function, we isolate it
as a coefficient in front of an appropriate tensor structure, namely
\begin{equation}
\frac{1}{\pi}ImT_{\mu \nu , \alpha \beta}{\epsilon}_{\alpha}
{\epsilon}^{*}_{\beta}
=\frac{i}{(pq)}{\epsilon}_{\mu \nu \lambda \sigma}
q_{\lambda}s_{\sigma}g_1^{\gamma}(x,Q^2,p^2).
\end{equation}
where $s_{\sigma}=i{\epsilon}_{\alpha \beta \gamma \sigma}
{\epsilon}_{\alpha}{\epsilon}^{*}_{\beta}p_{\gamma}$
and ${\epsilon}_{\alpha}$ is a photon polarization vector.
More precisely, for the purposes of this paper it is enough to
pick out the antisymmetric tensor
$(g_{\mu \alpha}g_{\nu \beta}-g_{\mu \beta}g_{\nu \alpha})$.

As a first step we have to find the contribution of the unit operator.
This result is well known since the "photon-photon fusion"
process was calculated even before the advent of QCD \cite{gin88}.
However, to reproduce unambiguously the spectral densities in the
dispersion representation for the structure function, which require
some intermediate result, such a calculation has to be performed over
again. We restrict ourselves to the scaling approximation, i.e. to
taking into account the first nonvanishing term in the expansion in
powers of $p^2/Q^2$, therefore limiting to the leading twist-2
contribution. The result is
\begin{eqnarray}
&&\!\!\!\!\!\!\!\!\!\!\!\!\!\!{g_1^{\gamma}}(x,Q^2,p^2)_{pert}\nonumber\\
&=&\frac{{\alpha}_{em}}{\pi}
N_cN_f \langle Q_q^4 \rangle
\left\{
\left[ 2-3x+[x^2-(1-x)^2]\int_{0}^{Q^2/x^2}
\frac{d{p'}^2 {p'}^2}{({p'}^2-{p}^2)^2}\right]-(1-x) \right\}\nonumber\\
&=&\frac{{\alpha}_{em}}{\pi}N_cN_f \langle Q_q^4 \rangle
[x^2-(1-x)^2]\left[ \ln\left( \frac{Q^2}{-p^2 x^2}\right)-2 \right],
\end{eqnarray}
where $\langle Q_q^4 \rangle=\frac{1}{N_f} \sum_{q}Q_q^4$ is an
average of the fourth powers of the quark charges. This result
is twice that represented by diagrams in fig. 2 due to the clockwise
and counter-clockwise directions of the internal quark lines, each
term in the curly brackets corresponding to graph (a) and (b),
respectively. The first line of this equation will be used in the
following to fix the parameters of the hadronic spectrum.

From all power corrections up to dimension eight we calculate only one
due to the gluon condensate $\langle \frac{{\alpha}_s}{\pi}G^2\rangle$.
This can be elucidated by the facts that the lowest dimension
quark condensate $\langle \bar{\psi} \psi \rangle$ cannot appear due to
chiral invariance as it is accompanied by the light quark mass which we
set equal to zero in all calculations. The contribution of the three-gluon
condensate $\langle g^3fG^3 \rangle$ is usually small. The dimension six
four-quark condensate can be omitted because its contribution is
proportional to the delta function --- $\delta(1-x)$ and turns out to
be beyond the scope of the method.

To simplify the calculation of the leading power correction,
it is convenient to use the fixed-point gauge for the background
gluon field $(x-x_0)_{\mu}B^a_{\mu}(x)=0$. We chose the
fixed point in the vertex of the hard photon emission $x_0=0$.
The quark propagator in this gauge up to the order $O(G^2)$ looks
like \cite{nov84}
\begin{eqnarray}
&&S(x,y)=\int \frac{d^4k}{(2\pi)^4} e^{-ik(x-y)}
\biggl\{ \frac{\not \! k}{k^2}
+g \widetilde G^a_{\alpha \beta}t^a
\frac{k_{\alpha}}{k^4}{\gamma}_{\beta}{\gamma}_5
+\frac{1}{2}g G^a_{\alpha \beta}t^a y_{\alpha}
\left[ \frac{{\gamma}_{\beta}}{k^2}
-2\frac{k_{\beta}\not \! k}{k^4} \right] \nonumber\\
&&\hspace{5cm}-\frac{\langle g^2G^2\rangle}{3^2 2^5}
\left[ \frac{y^2 \not \! k}{k^4}
+4\frac{\not \! k (ky)^2}{k^6}
-2\frac{\not \! y (ky)}{k^4} \right] \biggr\}.
\end{eqnarray}

A non-zero contribution comes in the leading twist from the diagrams
depicted in fig. 3 and each term in the curly brackets corresponds to the
diagrams (a), (b) and (c), respectively:
\begin{eqnarray}
&&\!\!\!\!\!\!\!\!\!\!\!\!\!
{g_1^{\gamma}}(x,Q^2,p^2)_{\langle \frac{{\alpha}_s}{\pi}G^2\rangle}
\nonumber\\
&=&\frac{{\alpha}_{em}}{\pi}N_cN_f \langle Q_q^4 \rangle \frac{1}{36}
\frac{{\pi}^2}{p^4} \langle \frac{{\alpha}_s}{\pi}G^2 \rangle
\left\{ \left[ \frac{8}{3}\frac{1}{x^2}-\frac{4}{3}\frac{1}{x} \right]
+\left[ \frac{8}{3}\frac{1}{x^2} \right]
+\left[ \frac{4}{3}\frac{1}{x} \right] \right\} \nonumber\\
&=&\frac{{\alpha}_{em}}{\pi}N_cN_f \langle Q_q^4 \rangle
\frac{4}{27}\frac{{\pi}^2}{p^4}
\langle \frac{{\alpha}_s}{\pi}G^2\rangle\frac{1}{x^2}.
\end{eqnarray}

Collecting all contributions we obtain the following structure
function for the off-shell polarized target photon:
\begin{eqnarray}
&&g_1^{\gamma}(x,Q^2,p^2)\nonumber\\
&&=\frac{{\alpha}_{em}}{\pi}N_cN_f \langle Q_q^4 \rangle
\left\{ [x^2-(1-x)^2]\left[ \ln\left( \frac{Q^2}{-p^2 x^2}\right)-2 \right]
+ \frac{4}{27} \frac{{\pi}^2 }{p^4}
\langle \frac{{\alpha}_s}{\pi}G^2\rangle\frac{1}{x^2} \right\}.\nonumber\\
\label{photstr}
\end{eqnarray}

The singularity in the gluon condensate contribution shows invalidity
of the method in the region of small $x$. The applicability region of
this formula, which accounts for the hadronic part, can be found by usual
requirement that the power corrections would not exceed $50\%$ of
the main perturbative term. Going as lower as to the point $-p^2
=m_{\rho}^2=0.6GeV^2$ ($\rho$-meson is the lowest prominent state in
the channel with the photon quantum numbers) in eq. (\ref{photstr})
the gluon condensate comprises less than $50\%$ only for $x\geq 0.6$.
The upper limit of permitted $x$-values will be estimated below. From
these facts it follows that the present approach could not be used for
the calculation of the moments of the structure function.

Now we are in a position to make some comments on the nonlocal condensates.

Introducing the nonlocal gluon condensate it is probably
impossible to get rid of singular behaviour of the corresponding
contribution to the structure function \cite{mik92}; the former
diminishes the degree of singularity by one power in the $x\sim 0$
region.

However, the singular $\delta$-type behaviour of the quark condensate
contribution can be smeared over the whole region of the momentum
fraction from zero to unity by introducing the concept of a nonlocal
quark condensate. Such an attempt was made in ref. \cite{bak94} for
the spin averaged structure function $F_2(x)$. But there are two
shortcomings in this paper. First, the authors claim that the
contribution due to the quark condensate improves considerably
the description of experimental data, though, there is a numerical
error in their answer: the coefficient in front of the nonlocal
vector quark condensate is three times smaller. The second fact
is connected with improper treatment of nonlocal objects. The
diagram with a nonlocal scalar condensate used in their paper does
not exist. The arguments are as follows. The philosophy of QCD sum rules
is the mutual relationship between the resonance and asymptotic parameters.
The leading asymptotic behaviour is given by perturbative QCD while
nonperturbative contributions enter through the power corrections
which die out for asymptotically large momenta squared of the problem
--- in the formal limit $-p^2$, $-\tau^2\rightarrow\infty$, where we
introduce nonzero $t$-channel momentum $\tau$ in order to treat the
factorization of small and large distances properly. The technique
is to extract the coefficient function that can be dealt perturbatively
--- the part of the diagram with hard momentum flow, while the soft lines
are parametrized via vacuum condensates that accumulate nonperturbative
information. In the diagram in ref. \cite{bak94} there is no
short-distance coefficient function. This can be traced by the fact
that in the local limit it becomes disconnected. To preserve the latter
property it is necessary to attach the gluon line connecting the upper
and lower parts (see fig. 4(a)). However, in the small $-\tau^2$
kinematics, actually, we are interested in the case of $-\tau^2=0$,
the diagram will behave as $\frac{1}{-\tau^2}$ for the local quark
condensates or something like $\frac{1}{-\tau^2+{\lambda}_q^2}$,
where $\frac{1}{{\lambda}_q}$ is a correlation length of quarks in
the vacuum, for the nonlocal one. For small $-\tau^2$ the use of
the bare gluon propagator is no longer reliable as the average
virtuality ${\lambda}_q^2=0.4GeV^2$ \cite{bel82} and turns outside
the perturbative domain. We expect that for low $-\tau^2$ the
nonperturbative effect does provide an appropriate infrared (IR)
cut-off: the perturbative behaviour $\frac{1}{-\tau^2}$ is
substituted by the mass of an appropriate meson $\frac{1}{m_R^2}$,
that is the "natural" scale for the boundary between the non- and
perturbative regions. In principle, this program could be traced
by reformation of the original OPE \cite{bal82} to the case when
the momentum $-\tau^2$ can be arbitrary small (even zero).
However, there is an essential difficulty as compared to the form
factor problem \cite{nes84}, namely, due to the presence of the
$t$-channel non-analyticities or singularities in each moment of
the structure function, we need an infinite number of parameters
to be found from additional sum rules which enter the model of the
bilocal power corrections. Obviously, this is an impossible task.
We will not go into details here referring the interested reader
to the paper \cite{bel96} where this question was studied for the
case of quark distribution in the pion. Therefore, limiting ourselves
to the central region in the Bjorken variable we discard these terms
completely. Of course, there are other diagrams (see fig. 4(b))
which are regular in $-\tau^2$ but we believe that the contribution
neglected could not affect seriously the final result.

\section{Polarized photon structure function}

We can use the analytical properties in $p^2$ \cite{bjo89} and represent
the structure function via a dispersion relation with respect to $p^2$
in terms of physical states
\begin{equation}
g_1^{\gamma}(x,Q^2,p^2)
=G_0(x)+\int_{0}^{\infty}d{p'}^2\frac{G_1(x,{p'}^2)}{({p'}^2-{p}^2)}
+\int_{0}^{\infty}d{p'_1}^2\int_{0}^{\infty}d{p'_2}^2
\frac{G_2(x,{p_1'}^2,{p_2'}^2)}{({p_1'}^2-{p}^2)({p_2'}^2-{p}^2)}.
\label{disp}
\end{equation}
To calculate the functions $G_i$, we use a standard technique in QCD
sum rules, viz., the simple model of a lowest resonance plus
continuum
\begin{eqnarray}
&&G_1(x,{p'}^2)=G^{(1)}_1(x)\delta({p'}^2-m^2_{\rho})
+G^{(2)}_1(x)\theta({p'}^2-{s_0}),\nonumber\\
&&G_2(x,{p_1'}^2,{p_2'}^2)\nonumber\\
&&=G^{(1)}_2(x)\delta({p_1'}^2-m^2_{\rho})\delta({p_2'}^2-m^2_{\rho})
+G^{(2)}_2(x)\theta({p_1'}^2-{s_0})\theta({p_2'}^2-{s_0}),
\end{eqnarray}
where $s_0=1.5GeV^2$ is a threshold value for the vector meson
channel and $m^2_{\rho}=0.6GeV^2$ is a ${\rho}$-meson mass squared.
In other words, we provide a natural cut-off on the transverse
momentum in the loop $\theta (k_{\perp}^2-x(1-x)s_0^2)$ attributing
the region of small $k_{\perp}^2$ to the nonperturbative contribution.
The continuum threshold $s_0$ has been found from the analysis
of two-point correlators for $\rho$-meson in the framework of
the QCD sum rules \cite{shi79}.

Requiring that at $-p^2\rightarrow \infty$ eq.(\ref{disp}) must coincide
with the bare quark loop, we obtain:
\begin{eqnarray}
&&G_0(x)=-\frac{{\alpha}_{em}}{\pi}N_cN_f \langle Q_q^4 \rangle
[x^2-(1-x)^2],\nonumber\\
&&G^{(2)}_1(x)=0,\nonumber\\
&&G^{(2)}_2(x)
=\frac{{\alpha}_{em}}{\pi}N_cN_f \langle Q_q^4 \rangle
[x^2-(1-x)^2]{p_1'}^2\delta({p_1'}^2-{p_2'}^2)\theta(Q^2/x^2-{p_1'}^2)
\end{eqnarray}
Substituting them back into the dispersion relation (\ref{disp}) and
expanding in the inverse powers of $p^2$ we can compare the resulting
expression with the QCD calculated $g_1^{\gamma}(x,Q^2,p^2)$
(\ref{photstr}) and fix the remaining unknown functions $G^{(1)}_1(x)$
and $G^{(1)}_2(x)$, namely:
\begin{eqnarray}
&&G^{(1)}_1(x)=0,\nonumber\\
&&G^{(1)}_2(x)=\frac{{\alpha}_{em}}{\pi}N_cN_f \langle Q_q^4 \rangle
\left[
[x^2-(1-x)^2]\frac{s_0^2}{2}
+\frac{4}{27}\pi^2 \langle \frac{{\alpha}_s}
{\pi}G^2\rangle\frac{1}{x^2}
\right].
\label{polro}
\end{eqnarray}

Finally, we collect all functions and perform the momentum integration
in eq.(\ref{disp}) keeping only the twist-2 contribution.
We come to the polarized virtual structure function which possesses the
correct analytical properties in the photon squared mass and
accounts for the hadronic part:
\begin{eqnarray}
&&g_1^{\gamma}(x,Q^2,p^2)
=\frac{{\alpha}_{em}}{\pi}N_cN_f \langle Q_q^4 \rangle\nonumber\\
&&\biggl\{ -[x^2-(1-x)^2]+[x^2-(1-x)^2]
\left[
\ln\left( \frac{Q^2}{x^2({s_0}-p^2)} \right)+\frac{p^2}{({s_0}-p^2)}
\right]\nonumber\\
&&+\frac{1}{(p^2-m^2_{\rho})^2}
\left[
[x^2-(1-x)^2]\frac{s_0^2}{2}
+\frac{4}{27}\pi^2 \langle \frac{{\alpha}_s}
{\pi}G^2\rangle\frac{1}{x^2}
\right]
\biggr\}.\label{final}
\end{eqnarray}

\section{Result and conclusion}

As we noticed above, the correction due to the gluon condensate
comprises no more than $50\%$ only for $x\geq 0.6$ at $-p^2
=m_{\rho}^2$. Therefore, for this $x$-values we can extrapolate
the polarized virtual structure function given by eq. (\ref{final})
to the point $p^2=0$. In the large-$x$ region, we could not trust
our model for hadronic spectrum. As was shown in ref. \cite{iof90}
as $x\to 1$ multihadron states exceed the contribution of single
$\rho$-meson. Moreover, since the function $G^{(1)}_2(x)$ given
by eq. (\ref{polro}) is related to the polarized structure function
of the meson\footnote{Eq. (\ref{polro}) corresponds to the local
duality relation for the structure function of the polarized
$\rho$-meson.} $\Delta f_{\rho}(x)$ it should reproduce $(1-x)$-behaviour
as $x\to 1$ governed by the quark counting rules. However, it is easy
to check that it does not take place. The upper limit of accepted
$x$-values may be estimated from he requirement that the created
hadron state should exceed considerably the $\rho$-meson mass
$M_X^2\gg m_{\rho}^2$. Then from the kinematical constraint
$x=1/(1+(M_X^2-m_{\rho}^2)/Q^2)$ we have the restriction $x<0.8$.
Therefore, the approach for calculating the polarized photon structure
function $g_1^{\gamma}(x)$ is valid in the interval $0.6\leq x<0.8$.
This situation is quite similar to that for calculating
the polarized proton structure functions in the QCD sum rules
framework where the range of permitted $x$-values is also very
narrow \cite{pol91}.

Note that from the point of view of the QCD sum rules an interpretation
of separate contributions to the structure function is different
from the na\"\i ve division of the latter in terms of the
perturbative (found from the box diagrams) and VMD parts which
inevitably contains double counting. In the language of the present
approach the part of the perturbative loop is dual to the lowest
resonance with corresponding quantum numbers and duality interval
$s_0$, while the remaining part of the loop corresponds to the higher
states contribution (the continuum). The resonance also receives
the contribution from the nonperturbative power corrections.

In fig. 5 we represent the result of calculation of the real
photon polarized structure function. The long- and short-dashed curves
correspond to the continuum and the vector-meson contributions to
the latter, respectively. They are represented by the first and
second terms in the expression (\ref{final}). The solid curve is
a full structure function (\ref{final}). We have used the standard
value for the gluon condensate $\langle \frac{{\alpha}_s}{\pi}G^2\rangle
=0.012GeV^4$ \cite{shi79}. From fig. 5. it is clearly seen that
the hadronic component obtained is very large.

Unfortunately, up to now there are no available experimental data on
the photon spin structure function $g_1^{\gamma}$ as its measurement
is at the limits of the possibilities of the polarized colliders.
Future measurements would provide an important insight into
the underlying dynamical effects associated with the polarized quark
and gluon content of the photon.

{\Large\bf Acknowledgments}

This work was supported by the International Science Foundation
under grant RFE300 and the Russian Foundation for Fundamental
Research under grant $N$ 93-02-3811.

\newpage
\pagestyle{empty}

\begin{figure}[h]
\unitlength=2.10pt
\special{em:linewidth 0.4pt}
\linethickness{0.4pt}
\begin{picture}(134.67,62.67)
\put(97.23,28.00){\oval(2.67,2.67)[lt]}
\put(97.23,30.66){\oval(2.67,2.67)[rb]}
\put(94.56,25.33){\oval(2.67,2.67)[lt]}
\put(94.56,28.00){\oval(2.67,2.67)[rb]}
\put(91.89,22.66){\oval(2.67,2.67)[lt]}
\put(91.89,25.33){\oval(2.67,2.67)[rb]}
\put(89.23,20.00){\oval(2.67,2.67)[lt]}
\put(89.23,22.66){\oval(2.67,2.67)[rb]}
\put(86.56,17.33){\oval(2.67,2.67)[lt]}
\put(86.56,20.00){\oval(2.67,2.67)[rb]}
\put(86.56,53.34){\oval(2.67,2.67)[lb]}
\put(86.56,50.68){\oval(2.67,2.67)[rt]}
\put(89.22,50.67){\oval(2.67,2.67)[lb]}
\put(89.23,48.01){\oval(2.67,2.67)[rt]}
\put(91.89,48.01){\oval(2.67,2.67)[lb]}
\put(91.90,45.34){\oval(2.67,2.67)[rt]}
\put(94.56,45.34){\oval(2.67,2.67)[lb]}
\put(94.56,42.68){\oval(2.67,2.67)[rt]}
\put(97.22,42.67){\oval(2.67,2.67)[lb]}
\put(97.23,40.01){\oval(2.67,2.67)[rt]}
\put(85.66,26.33){\makebox(0,0)[cc]{$p$}}
\put(85.66,44.34){\makebox(0,0)[cc]{$q$}}
\put(92.50,24.18){\vector(1,1){1.11}}
\put(92.54,46.65){\vector(1,-1){1.11}}
\put(103.33,35.33){\circle{14.00}}
\put(70.33,17.33){\line(1,0){15.00}}
\put(85.33,17.33){\line(3,-2){14.00}}
\put(99.33,62.67){\line(-3,-2){14.00}}
\put(85.33,53.33){\line(-1,0){15.00}}
\put(76.33,53.33){\vector(1,0){1.00}}
\put(91.66,57.67){\vector(3,2){1.00}}
\put(76.33,17.33){\vector(1,0){1.00}}
\put(90.99,13.67){\vector(4,-3){1.67}}
\put(110.00,35.33){\line(1,0){15.00}}
\put(109.67,37.33){\line(6,1){15.33}}
\put(109.67,33.33){\line(5,-1){15.33}}
\put(70.00,58.00){\makebox(0,0)[cc]{$e^{\pm}$}}
\put(70.00,12.67){\makebox(0,0)[cc]{$e^-$}}
\put(134.67,35.33){\makebox(0,0)[cc]{$X$}}
\end{picture}
\caption{ The kinematics of the two-photon reaction
$e^{\pm}e^-\to e^{\pm}e^-X$.}
\end{figure}
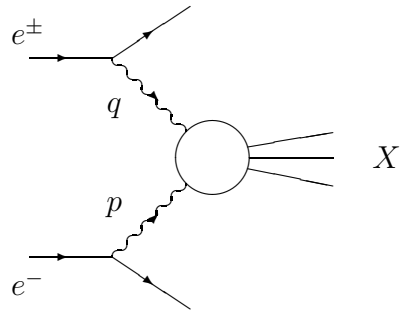

\vspace{1cm}

\begin{figure}[h]
\unitlength=2.10pt
\special{em:linewidth 0.4pt}
\linethickness{0.4pt}
\begin{picture}(169.90,60.33)
\put(128.57,25.67){\vector(0,1){11.00}}
\put(128.57,36.67){\line(0,1){9.00}}
\put(78.57,45.67){\vector(0,-1){11.00}}
\put(78.57,34.67){\line(0,-1){9.00}}
\put(58.57,25.67){\vector(0,1){11.00}}
\put(58.57,36.67){\line(0,1){9.00}}
\put(58.57,45.67){\vector(1,0){11.00}}
\put(69.57,45.67){\line(1,0){9.00}}
\put(128.57,45.67){\vector(1,0){11.00}}
\put(139.57,45.67){\line(1,0){9.00}}
\put(148.57,25.66){\vector(-1,0){11.00}}
\put(78.57,25.66){\vector(-1,0){11.00}}
\put(67.57,25.66){\line(-1,0){9.00}}
\put(137.57,25.66){\line(-1,0){9.00}}
\put(90.57,56.34){\oval(2.67,2.67)[lt]}
\put(90.57,59.00){\oval(2.67,2.67)[rb]}
\put(87.90,53.67){\oval(2.67,2.67)[lt]}
\put(87.90,56.34){\oval(2.67,2.67)[rb]}
\put(85.23,51.00){\oval(2.67,2.67)[lt]}
\put(85.23,53.67){\oval(2.67,2.67)[rb]}
\put(82.57,48.34){\oval(2.67,2.67)[lt]}
\put(82.57,51.00){\oval(2.67,2.67)[rb]}
\put(79.90,45.67){\oval(2.67,2.67)[lt]}
\put(79.90,48.34){\oval(2.67,2.67)[rb]}
\put(57.24,23.00){\oval(2.67,2.67)[lt]}
\put(57.24,25.66){\oval(2.67,2.67)[rb]}
\put(54.57,20.33){\oval(2.67,2.67)[lt]}
\put(54.57,23.00){\oval(2.67,2.67)[rb]}
\put(51.90,17.66){\oval(2.67,2.67)[lt]}
\put(51.90,20.33){\oval(2.67,2.67)[rb]}
\put(49.24,15.00){\oval(2.67,2.67)[lt]}
\put(49.24,17.66){\oval(2.67,2.67)[rb]}
\put(46.57,12.33){\oval(2.67,2.67)[lt]}
\put(46.57,15.00){\oval(2.67,2.67)[rb]}
\put(127.24,23.01){\oval(2.67,2.67)[lt]}
\put(127.24,25.67){\oval(2.67,2.67)[rb]}
\put(124.57,20.34){\oval(2.67,2.67)[lt]}
\put(124.57,23.01){\oval(2.67,2.67)[rb]}
\put(121.90,17.67){\oval(2.67,2.67)[lt]}
\put(121.90,20.34){\oval(2.67,2.67)[rb]}
\put(119.24,15.01){\oval(2.67,2.67)[lt]}
\put(119.24,17.67){\oval(2.67,2.67)[rb]}
\put(116.57,12.34){\oval(2.67,2.67)[lt]}
\put(116.57,15.01){\oval(2.67,2.67)[rb]}
\put(168.57,44.34){\oval(2.67,2.67)[lt]}
\put(168.57,47.00){\oval(2.67,2.67)[rb]}
\put(165.90,41.67){\oval(2.67,2.67)[lt]}
\put(165.90,44.34){\oval(2.67,2.67)[rb]}
\put(163.23,39.00){\oval(2.67,2.67)[lt]}
\put(163.23,41.67){\oval(2.67,2.67)[rb]}
\put(160.57,36.34){\oval(2.67,2.67)[lt]}
\put(160.57,39.00){\oval(2.67,2.67)[rb]}
\put(157.90,33.67){\oval(2.67,2.67)[lt]}
\put(157.90,36.34){\oval(2.67,2.67)[rb]}
\put(155.23,31.00){\oval(2.67,2.67)[lt]}
\put(155.23,33.67){\oval(2.67,2.67)[rb]}
\put(152.57,28.34){\oval(2.67,2.67)[lt]}
\put(152.57,31.00){\oval(2.67,2.67)[rb]}
\put(149.90,25.67){\oval(2.67,2.67)[lt]}
\put(149.90,28.34){\oval(2.67,2.67)[rb]}
\put(148.57,45.67){\vector(0,-1){11.00}}
\put(148.57,34.67){\line(0,-1){9.00}}
\put(46.57,59.00){\oval(2.67,2.67)[lb]}
\put(46.57,56.34){\oval(2.67,2.67)[rt]}
\put(49.23,56.33){\oval(2.67,2.67)[lb]}
\put(49.24,53.67){\oval(2.67,2.67)[rt]}
\put(51.90,53.67){\oval(2.67,2.67)[lb]}
\put(51.91,51.00){\oval(2.67,2.67)[rt]}
\put(54.57,51.00){\oval(2.67,2.67)[lb]}
\put(54.57,48.34){\oval(2.67,2.67)[rt]}
\put(57.23,48.33){\oval(2.67,2.67)[lb]}
\put(57.24,45.67){\oval(2.67,2.67)[rt]}
\put(79.90,25.67){\oval(2.67,2.67)[lb]}
\put(79.91,23.00){\oval(2.67,2.67)[rt]}
\put(82.57,23.00){\oval(2.67,2.67)[lb]}
\put(82.57,20.34){\oval(2.67,2.67)[rt]}
\put(85.23,20.33){\oval(2.67,2.67)[lb]}
\put(85.24,17.67){\oval(2.67,2.67)[rt]}
\put(87.90,17.67){\oval(2.67,2.67)[lb]}
\put(87.91,15.00){\oval(2.67,2.67)[rt]}
\put(90.57,15.00){\oval(2.67,2.67)[lb]}
\put(90.57,12.34){\oval(2.67,2.67)[rt]}
\put(116.57,59.00){\oval(2.67,2.67)[lb]}
\put(116.57,56.34){\oval(2.67,2.67)[rt]}
\put(119.23,56.33){\oval(2.67,2.67)[lb]}
\put(119.24,53.67){\oval(2.67,2.67)[rt]}
\put(121.90,53.67){\oval(2.67,2.67)[lb]}
\put(121.91,51.00){\oval(2.67,2.67)[rt]}
\put(124.57,51.00){\oval(2.67,2.67)[lb]}
\put(124.57,48.34){\oval(2.67,2.67)[rt]}
\put(127.23,48.33){\oval(2.67,2.67)[lb]}
\put(127.24,45.67){\oval(2.67,2.67)[rt]}
\put(149.90,45.67){\oval(2.67,2.67)[lb]}
\put(149.91,43.00){\oval(2.67,2.67)[rt]}
\put(152.57,43.00){\oval(2.67,2.67)[lb]}
\put(152.57,40.34){\oval(2.67,2.67)[rt]}
\put(155.23,40.33){\oval(2.67,2.67)[lb]}
\put(155.24,37.67){\oval(2.67,2.67)[rt]}
\put(157.90,37.67){\oval(2.67,2.67)[lb]}
\put(157.91,35.00){\oval(2.67,2.67)[rt]}
\put(160.57,35.00){\oval(2.67,2.67)[lb]}
\put(160.57,32.34){\oval(2.67,2.67)[rt]}
\put(163.23,32.33){\oval(2.67,2.67)[lb]}
\put(163.24,29.67){\oval(2.67,2.67)[rt]}
\put(165.90,29.67){\oval(2.67,2.67)[lb]}
\put(165.91,27.00){\oval(2.67,2.67)[rt]}
\put(168.57,27.00){\oval(2.67,2.67)[lb]}
\put(168.57,24.34){\oval(2.67,2.67)[rt]}
\put(68.57,4.67){\makebox(0,0)[cc]{(a)}}
\put(138.57,4.67){\makebox(0,0)[cc]{(b)}}
\put(53.67,59.33){\makebox(0,0)[cc]{$\mu$}}
\put(53.67,12.00){\makebox(0,0)[cc]{$\alpha$}}
\put(83.67,12.00){\makebox(0,0)[cc]{$\beta$}}
\put(83.67,59.33){\makebox(0,0)[cc]{$\nu$}}
\put(45.67,21.33){\makebox(0,0)[cc]{$p$}}
\put(45.67,50.00){\makebox(0,0)[cc]{$q$}}
\put(123.67,59.33){\makebox(0,0)[cc]{$\mu$}}
\put(123.67,12.00){\makebox(0,0)[cc]{$\alpha$}}
\put(164.78,19.41){\makebox(0,0)[cc]{$\beta$}}
\put(164.78,51.92){\makebox(0,0)[cc]{$\nu$}}
\put(162.22,39.63){\vector(1,1){1.11}}
\put(164.07,30.75){\vector(1,-1){1.11}}
\put(122.51,19.18){\vector(1,1){1.11}}
\put(122.55,52.31){\vector(1,-1){1.11}}
\put(84.22,51.89){\vector(1,1){1.11}}
\put(84.44,19.53){\vector(1,-1){1.11}}
\put(52.51,19.18){\vector(1,1){1.11}}
\put(52.55,52.31){\vector(1,-1){1.11}}
\end{picture}
\caption{ Perturbative diagrams for the unit
operator in the operator expansion.}
\end{figure}
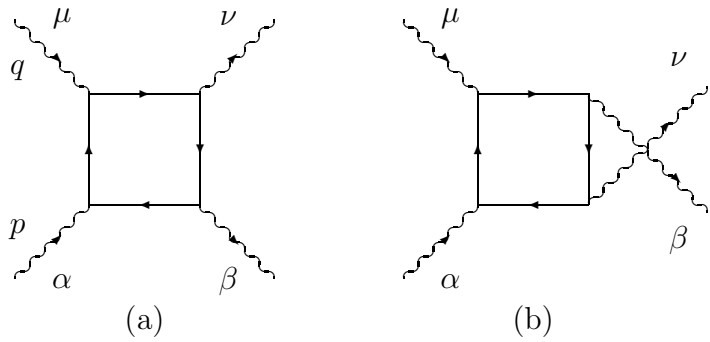

\vspace{1cm}

\begin{figure}[h]
\unitlength=2.10pt
\special{em:linewidth 0.4pt}
\linethickness{0.4pt}
\begin{picture}(191.02,62.05)
\put(37.73,47.29){\vector(1,0){11.00}}
\put(48.73,47.29){\line(1,0){9.00}}
\put(57.73,27.28){\vector(-1,0){11.00}}
\put(46.73,27.28){\line(-1,0){9.00}}
\
\
\put(117.73,47.29){\vector(0,-1){11.00}}
\put(117.73,36.29){\line(0,-1){9.00}}
\put(97.73,47.29){\vector(1,0){11.00}}
\put(108.73,47.29){\line(1,0){9.00}}
\put(177.70,47.23){\vector(0,-1){11.00}}
\put(177.70,36.23){\line(0,-1){9.00}}
\put(157.70,27.23){\vector(0,1){11.00}}
\put(157.70,38.23){\line(0,1){9.00}}
\put(157.70,47.23){\vector(1,0){11.00}}
\put(168.70,47.23){\line(1,0){9.00}}
\put(57.73,47.29){\vector(0,-1){6.67}}
\put(57.73,40.62){\vector(0,-1){10.00}}
\put(57.73,30.62){\line(0,-1){3.33}}
\put(37.73,27.29){\vector(0,1){6.67}}
\put(37.73,33.95){\vector(0,1){10.00}}
\put(37.73,43.95){\line(0,1){3.33}}
\put(59.57,36.66){\oval(3.67,3.67)[t]}
\put(60.57,36.49){\oval(1.67,2.67)[b]}
\put(61.57,36.66){\oval(3.67,3.67)[t]}
\put(62.57,36.49){\oval(1.67,2.67)[b]}
\put(63.57,36.66){\oval(3.67,3.67)[t]}
\put(64.57,36.49){\oval(1.67,2.67)[b]}
\put(65.57,36.66){\oval(3.67,3.67)[t]}
\put(66.57,36.49){\oval(1.67,2.67)[b]}
\put(67.57,36.66){\oval(3.67,3.67)[t]}
\put(68.57,36.49){\oval(1.67,2.67)[b]}
\put(69.57,36.66){\oval(3.67,3.67)[t]}
\put(25.90,36.66){\oval(3.67,3.67)[t]}
\put(26.90,36.49){\oval(1.67,2.67)[b]}
\put(27.90,36.66){\oval(3.67,3.67)[t]}
\put(28.90,36.49){\oval(1.67,2.67)[b]}
\put(29.90,36.66){\oval(3.67,3.67)[t]}
\put(30.90,36.49){\oval(1.67,2.67)[b]}
\put(31.90,36.66){\oval(3.67,3.67)[t]}
\put(32.90,36.49){\oval(1.67,2.67)[b]}
\put(33.90,36.66){\oval(3.67,3.67)[t]}
\put(34.90,36.49){\oval(1.67,2.67)[b]}
\put(35.90,36.66){\oval(3.67,3.67)[t]}
\put(97.72,27.29){\vector(0,1){6.67}}
\put(97.72,33.95){\vector(0,1){10.00}}
\put(97.72,43.95){\line(0,1){3.33}}
\put(85.89,36.66){\oval(3.67,3.67)[t]}
\put(86.89,36.49){\oval(1.67,2.67)[b]}
\put(87.89,36.66){\oval(3.67,3.67)[t]}
\put(88.89,36.49){\oval(1.67,2.67)[b]}
\put(89.89,36.66){\oval(3.67,3.67)[t]}
\put(90.89,36.49){\oval(1.67,2.67)[b]}
\put(91.89,36.66){\oval(3.67,3.67)[t]}
\put(92.89,36.49){\oval(1.67,2.67)[b]}
\put(93.89,36.66){\oval(3.67,3.67)[t]}
\put(94.89,36.49){\oval(1.67,2.67)[b]}
\put(95.89,36.66){\oval(3.67,3.67)[t]}
\put(117.73,27.29){\vector(-1,0){6.67}}
\put(111.06,27.29){\vector(-1,0){10.00}}
\put(101.06,27.29){\line(-1,0){3.33}}
\put(107.69,25.62){\oval(3.01,3.01)[l]}
\put(107.69,17.62){\oval(3.01,3.01)[l]}
\put(108.63,18.68){\oval(1.67,1.00)[r]}
\put(107.69,19.62){\oval(3.01,3.01)[l]}
\put(108.63,20.68){\oval(1.67,1.00)[r]}
\put(107.69,21.62){\oval(3.01,3.01)[l]}
\put(108.63,22.68){\oval(1.67,1.00)[r]}
\put(107.69,23.62){\oval(3.01,3.01)[l]}
\put(108.63,24.68){\oval(1.67,1.00)[r]}
\put(107.69,15.62){\oval(3.01,3.01)[l]}
\put(108.63,16.68){\oval(1.67,1.00)[r]}
\put(174.51,27.11){\vector(-1,0){7.67}}
\put(177.51,27.11){\vector(-1,0){3.33}}
\put(166.84,27.11){\vector(-1,0){7.67}}
\put(159.18,27.11){\line(-1,0){1.33}}
\put(163.82,25.44){\oval(3.01,3.01)[l]}
\put(163.82,17.44){\oval(3.01,3.01)[l]}
\put(164.76,18.50){\oval(1.67,1.00)[r]}
\put(163.82,19.44){\oval(3.01,3.01)[l]}
\put(164.76,20.50){\oval(1.67,1.00)[r]}
\put(163.82,21.44){\oval(3.01,3.01)[l]}
\put(164.76,22.50){\oval(1.67,1.00)[r]}
\put(163.82,23.44){\oval(3.01,3.01)[l]}
\put(164.76,24.50){\oval(1.67,1.00)[r]}
\put(163.82,15.44){\oval(3.01,3.01)[l]}
\put(164.76,16.50){\oval(1.67,1.00)[r]}
\put(171.48,25.44){\oval(3.01,3.01)[l]}
\put(171.48,17.44){\oval(3.01,3.01)[l]}
\put(172.42,18.50){\oval(1.67,1.00)[r]}
\put(171.48,19.44){\oval(3.01,3.01)[l]}
\put(172.42,20.50){\oval(1.67,1.00)[r]}
\put(171.48,21.44){\oval(3.01,3.01)[l]}
\put(172.42,22.50){\oval(1.67,1.00)[r]}
\put(171.48,23.44){\oval(3.01,3.01)[l]}
\put(172.42,24.50){\oval(1.67,1.00)[r]}
\put(171.48,15.44){\oval(3.01,3.01)[l]}
\put(172.42,16.50){\oval(1.67,1.00)[r]}
\put(24.06,36.51){\makebox(0,0)[cc]{$\times$}}
\put(71.18,36.51){\makebox(0,0)[cc]{$\times$}}
\put(84.06,36.51){\makebox(0,0)[cc]{$\times$}}
\put(108.18,13.40){\makebox(0,0)[cc]{$\times$}}
\put(164.29,12.99){\makebox(0,0)[cc]{$\times$}}
\put(172.06,12.99){\makebox(0,0)[cc]{$\times$}}
\put(47.62,4.29){\makebox(0,0)[cc]{(a)}}
\put(107.62,4.29){\makebox(0,0)[cc]{(b)}}
\put(167.62,4.29){\makebox(0,0)[cc]{(c)}}
\put(37.78,37.04){\circle*{1.48}}
\put(57.78,37.04){\circle*{1.48}}
\put(97.78,37.04){\circle*{1.48}}
\put(108.15,27.41){\circle*{1.48}}
\put(164.44,27.04){\circle*{1.48}}
\put(172.22,27.04){\circle*{1.48}}
\put(189.69,58.06){\oval(2.67,2.67)[lt]}
\put(189.69,60.72){\oval(2.67,2.67)[rb]}
\put(187.02,55.39){\oval(2.67,2.67)[lt]}
\put(187.02,58.06){\oval(2.67,2.67)[rb]}
\put(184.35,52.72){\oval(2.67,2.67)[lt]}
\put(184.35,55.39){\oval(2.67,2.67)[rb]}
\put(181.69,50.06){\oval(2.67,2.67)[lt]}
\put(181.69,52.72){\oval(2.67,2.67)[rb]}
\put(179.02,47.39){\oval(2.67,2.67)[lt]}
\put(179.02,50.06){\oval(2.67,2.67)[rb]}
\put(156.36,24.72){\oval(2.67,2.67)[lt]}
\put(156.36,27.38){\oval(2.67,2.67)[rb]}
\put(153.69,22.05){\oval(2.67,2.67)[lt]}
\put(153.69,24.72){\oval(2.67,2.67)[rb]}
\put(151.02,19.38){\oval(2.67,2.67)[lt]}
\put(151.02,22.05){\oval(2.67,2.67)[rb]}
\put(148.36,16.72){\oval(2.67,2.67)[lt]}
\put(148.36,19.38){\oval(2.67,2.67)[rb]}
\put(145.69,14.05){\oval(2.67,2.67)[lt]}
\put(145.69,16.72){\oval(2.67,2.67)[rb]}
\put(145.69,60.72){\oval(2.67,2.67)[lb]}
\put(145.69,58.06){\oval(2.67,2.67)[rt]}
\put(148.35,58.05){\oval(2.67,2.67)[lb]}
\put(148.36,55.39){\oval(2.67,2.67)[rt]}
\put(151.02,55.39){\oval(2.67,2.67)[lb]}
\put(151.03,52.72){\oval(2.67,2.67)[rt]}
\put(153.69,52.72){\oval(2.67,2.67)[lb]}
\put(153.69,50.06){\oval(2.67,2.67)[rt]}
\put(156.35,50.05){\oval(2.67,2.67)[lb]}
\put(156.36,47.39){\oval(2.67,2.67)[rt]}
\put(179.02,27.39){\oval(2.67,2.67)[lb]}
\put(179.03,24.72){\oval(2.67,2.67)[rt]}
\put(181.69,24.72){\oval(2.67,2.67)[lb]}
\put(181.69,22.06){\oval(2.67,2.67)[rt]}
\put(184.35,22.05){\oval(2.67,2.67)[lb]}
\put(184.36,19.39){\oval(2.67,2.67)[rt]}
\put(187.02,19.39){\oval(2.67,2.67)[lb]}
\put(187.03,16.72){\oval(2.67,2.67)[rt]}
\put(189.69,16.72){\oval(2.67,2.67)[lb]}
\put(189.69,14.06){\oval(2.67,2.67)[rt]}
\put(183.34,53.61){\vector(1,1){1.11}}
\put(183.56,21.25){\vector(1,-1){1.11}}
\put(151.63,20.90){\vector(1,1){1.11}}
\put(151.67,54.03){\vector(1,-1){1.11}}
\put(69.78,58.06){\oval(2.67,2.67)[lt]}
\put(69.78,60.72){\oval(2.67,2.67)[rb]}
\put(67.11,55.39){\oval(2.67,2.67)[lt]}
\put(67.11,58.06){\oval(2.67,2.67)[rb]}
\put(64.44,52.72){\oval(2.67,2.67)[lt]}
\put(64.44,55.39){\oval(2.67,2.67)[rb]}
\put(61.78,50.06){\oval(2.67,2.67)[lt]}
\put(61.78,52.72){\oval(2.67,2.67)[rb]}
\put(59.11,47.39){\oval(2.67,2.67)[lt]}
\put(59.11,50.06){\oval(2.67,2.67)[rb]}
\put(36.45,24.72){\oval(2.67,2.67)[lt]}
\put(36.45,27.38){\oval(2.67,2.67)[rb]}
\put(33.78,22.05){\oval(2.67,2.67)[lt]}
\put(33.78,24.72){\oval(2.67,2.67)[rb]}
\put(31.11,19.38){\oval(2.67,2.67)[lt]}
\put(31.11,22.05){\oval(2.67,2.67)[rb]}
\put(28.45,16.72){\oval(2.67,2.67)[lt]}
\put(28.45,19.38){\oval(2.67,2.67)[rb]}
\put(25.78,14.05){\oval(2.67,2.67)[lt]}
\put(25.78,16.72){\oval(2.67,2.67)[rb]}
\put(25.78,60.72){\oval(2.67,2.67)[lb]}
\put(25.78,58.06){\oval(2.67,2.67)[rt]}
\put(28.44,58.05){\oval(2.67,2.67)[lb]}
\put(28.45,55.39){\oval(2.67,2.67)[rt]}
\put(31.11,55.39){\oval(2.67,2.67)[lb]}
\put(31.12,52.72){\oval(2.67,2.67)[rt]}
\put(33.78,52.72){\oval(2.67,2.67)[lb]}
\put(33.78,50.06){\oval(2.67,2.67)[rt]}
\put(36.44,50.05){\oval(2.67,2.67)[lb]}
\put(36.45,47.39){\oval(2.67,2.67)[rt]}
\put(59.11,27.39){\oval(2.67,2.67)[lb]}
\put(59.12,24.72){\oval(2.67,2.67)[rt]}
\put(61.78,24.72){\oval(2.67,2.67)[lb]}
\put(61.78,22.06){\oval(2.67,2.67)[rt]}
\put(64.44,22.05){\oval(2.67,2.67)[lb]}
\put(64.45,19.39){\oval(2.67,2.67)[rt]}
\put(67.11,19.39){\oval(2.67,2.67)[lb]}
\put(67.12,16.72){\oval(2.67,2.67)[rt]}
\put(69.78,16.72){\oval(2.67,2.67)[lb]}
\put(69.78,14.06){\oval(2.67,2.67)[rt]}
\put(63.43,53.61){\vector(1,1){1.11}}
\put(63.65,21.25){\vector(1,-1){1.11}}
\put(31.72,20.90){\vector(1,1){1.11}}
\put(31.76,54.03){\vector(1,-1){1.11}}
\put(129.61,57.97){\oval(2.67,2.67)[lt]}
\put(129.61,60.63){\oval(2.67,2.67)[rb]}
\put(126.94,55.30){\oval(2.67,2.67)[lt]}
\put(126.94,57.97){\oval(2.67,2.67)[rb]}
\put(124.27,52.63){\oval(2.67,2.67)[lt]}
\put(124.27,55.30){\oval(2.67,2.67)[rb]}
\put(121.61,49.97){\oval(2.67,2.67)[lt]}
\put(121.61,52.63){\oval(2.67,2.67)[rb]}
\put(118.94,47.30){\oval(2.67,2.67)[lt]}
\put(118.94,49.97){\oval(2.67,2.67)[rb]}
\put(96.28,24.63){\oval(2.67,2.67)[lt]}
\put(96.28,27.29){\oval(2.67,2.67)[rb]}
\put(93.61,21.96){\oval(2.67,2.67)[lt]}
\put(93.61,24.63){\oval(2.67,2.67)[rb]}
\put(90.94,19.29){\oval(2.67,2.67)[lt]}
\put(90.94,21.96){\oval(2.67,2.67)[rb]}
\put(88.28,16.63){\oval(2.67,2.67)[lt]}
\put(88.28,19.29){\oval(2.67,2.67)[rb]}
\put(85.61,13.96){\oval(2.67,2.67)[lt]}
\put(85.61,16.63){\oval(2.67,2.67)[rb]}
\put(85.61,60.63){\oval(2.67,2.67)[lb]}
\put(85.61,57.97){\oval(2.67,2.67)[rt]}
\put(88.27,57.96){\oval(2.67,2.67)[lb]}
\put(88.28,55.30){\oval(2.67,2.67)[rt]}
\put(90.94,55.30){\oval(2.67,2.67)[lb]}
\put(90.95,52.63){\oval(2.67,2.67)[rt]}
\put(93.61,52.63){\oval(2.67,2.67)[lb]}
\put(93.61,49.97){\oval(2.67,2.67)[rt]}
\put(96.27,49.96){\oval(2.67,2.67)[lb]}
\put(96.28,47.30){\oval(2.67,2.67)[rt]}
\put(118.94,27.30){\oval(2.67,2.67)[lb]}
\put(118.95,24.63){\oval(2.67,2.67)[rt]}
\put(121.61,24.63){\oval(2.67,2.67)[lb]}
\put(121.61,21.97){\oval(2.67,2.67)[rt]}
\put(124.27,21.96){\oval(2.67,2.67)[lb]}
\put(124.28,19.30){\oval(2.67,2.67)[rt]}
\put(126.94,19.30){\oval(2.67,2.67)[lb]}
\put(126.95,16.63){\oval(2.67,2.67)[rt]}
\put(129.61,16.63){\oval(2.67,2.67)[lb]}
\put(129.61,13.97){\oval(2.67,2.67)[rt]}
\put(123.26,53.52){\vector(1,1){1.11}}
\put(123.48,21.16){\vector(1,-1){1.11}}
\put(91.55,20.81){\vector(1,1){1.11}}
\put(91.59,53.94){\vector(1,-1){1.11}}
\end{picture}
\caption{ Gluon condensate contribution to the imaginary part
of the forward $\gamma\gamma$-scattering amplitude.}
\end{figure}
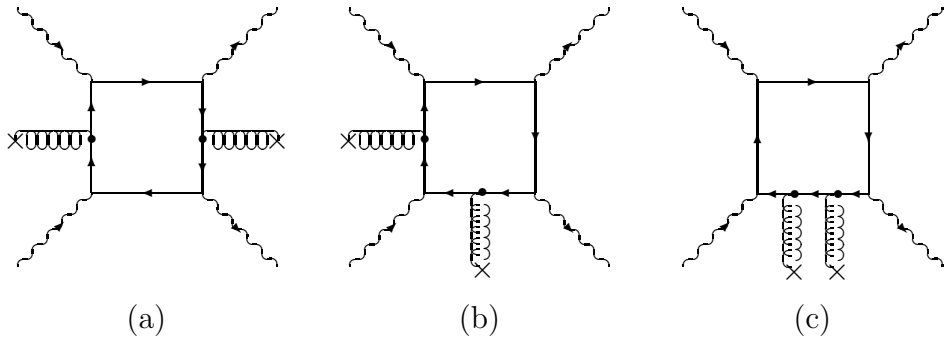

\newpage
\pagestyle{empty}

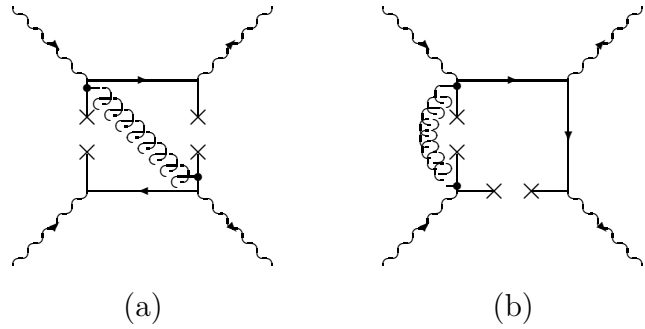
\begin{figure}[h]
\unitlength=2.10pt
\special{em:linewidth 0.4pt}
\linethickness{0.4pt}
\begin{picture}(154.94,60.30)
\
\
\put(55.04,45.67){\vector(1,0){11.00}}
\put(66.04,45.67){\line(1,0){9.00}}
\put(75.04,25.66){\vector(-1,0){11.00}}
\put(64.04,25.66){\line(-1,0){9.00}}
\put(65.04,4.67){\makebox(0,0)[cc]{(a)}}
\put(59.71,40.85){\oval(3.00,3.00)[r]}
\put(56.86,41.85){\oval(1.33,1.00)[l]}
\put(57.20,42.35){\line(1,0){2.00}}
\put(58.87,39.85){\oval(1.33,1.00)[l]}
\put(61.70,38.85){\oval(3.00,3.00)[r]}
\put(63.71,36.85){\oval(3.00,3.00)[r]}
\put(60.86,37.85){\oval(1.33,1.00)[l]}
\put(61.20,38.35){\line(1,0){2.00}}
\put(62.87,35.85){\oval(1.33,1.00)[l]}
\put(59.20,40.35){\line(1,0){2.00}}
\put(65.70,34.85){\oval(3.00,3.00)[r]}
\put(67.71,32.85){\oval(3.00,3.00)[r]}
\put(64.86,33.85){\oval(1.33,1.00)[l]}
\put(65.20,34.35){\line(1,0){2.00}}
\put(66.87,31.85){\oval(1.33,1.00)[l]}
\put(63.20,36.35){\line(1,0){2.00}}
\put(69.70,30.85){\oval(3.00,3.00)[r]}
\put(71.71,28.85){\oval(3.00,3.00)[r]}
\put(68.86,29.85){\oval(1.33,1.00)[l]}
\put(69.20,30.35){\line(1,0){2.00}}
\put(70.87,27.85){\oval(1.33,1.00)[l]}
\put(67.20,32.35){\line(1,0){2.00}}
\put(71.30,28.33){\line(1,0){3.67}}
\put(57.71,42.85){\oval(3.00,3.00)[r]}
\put(55.20,44.35){\line(1,0){2.00}}
\put(54.97,44.33){\circle*{1.33}}
\put(74.97,28.33){\circle*{1.33}}
\put(54.97,45.67){\line(0,-1){6.67}}
\put(54.97,25.67){\line(0,1){6.67}}
\put(74.97,45.67){\line(0,-1){6.67}}
\put(74.97,25.67){\line(0,1){6.67}}
\put(54.97,39.00){\makebox(0,0)[cc]{$\times$}}
\put(54.97,32.67){\makebox(0,0)[cc]{$\times$}}
\put(74.97,32.67){\makebox(0,0)[cc]{$\times$}}
\put(74.97,39.00){\makebox(0,0)[cc]{$\times$}}
\put(121.71,45.67){\vector(1,0){11.00}}
\put(132.71,45.67){\line(1,0){9.00}}
\put(131.71,4.67){\makebox(0,0)[cc]{(b)}}
\put(121.64,45.67){\line(0,-1){6.67}}
\put(121.64,25.67){\line(0,1){6.67}}
\put(121.64,39.00){\makebox(0,0)[cc]{$\times$}}
\put(121.64,32.67){\makebox(0,0)[cc]{$\times$}}
\put(141.64,45.67){\line(0,-1){20.00}}
\put(141.64,35.67){\vector(0,-1){1.00}}
\put(121.64,25.67){\line(1,0){6.67}}
\put(141.64,25.67){\line(-1,0){6.67}}
\put(134.97,25.67){\makebox(0,0)[cc]{$\times$}}
\put(128.30,25.67){\makebox(0,0)[cc]{$\times$}}
\put(116.53,37.15){\oval(3.01,3.01)[l]}
\put(116.86,39.15){\oval(3.01,3.01)[l]}
\put(117.86,41.15){\oval(3.01,3.01)[l]}
\put(119.20,43.15){\oval(3.01,3.01)[l]}
\put(117.47,38.21){\oval(1.67,1.00)[r]}
\put(118.47,40.21){\oval(1.67,1.00)[r]}
\put(119.81,42.21){\oval(1.67,1.00)[r]}
\put(119.19,29.16){\oval(3.01,3.01)[l]}
\put(117.86,31.16){\oval(3.01,3.01)[l]}
\put(116.86,33.16){\oval(3.01,3.01)[l]}
\put(116.53,35.16){\oval(3.01,3.01)[l]}
\put(119.80,30.22){\oval(1.67,1.00)[r]}
\put(118.47,32.22){\oval(1.67,1.00)[r]}
\put(117.47,34.22){\oval(1.67,1.00)[r]}
\put(117.14,36.22){\oval(1.67,1.00)[r]}
\put(116.98,40.73){\line(1,0){1.33}}
\put(118.64,42.73){\line(1,0){1.00}}
\put(118.31,28.73){\line(1,0){1.33}}
\put(117.31,30.73){\line(1,0){1.00}}
\put(119.30,44.67){\line(1,0){2.33}}
\put(121.64,26.67){\line(-1,0){2.00}}
\put(121.64,44.67){\circle*{1.33}}
\put(121.64,26.67){\circle*{1.33}}
\put(153.61,56.31){\oval(2.67,2.67)[lt]}
\put(153.61,58.97){\oval(2.67,2.67)[rb]}
\put(150.94,53.64){\oval(2.67,2.67)[lt]}
\put(150.94,56.31){\oval(2.67,2.67)[rb]}
\put(148.27,50.97){\oval(2.67,2.67)[lt]}
\put(148.27,53.64){\oval(2.67,2.67)[rb]}
\put(145.61,48.31){\oval(2.67,2.67)[lt]}
\put(145.61,50.97){\oval(2.67,2.67)[rb]}
\put(142.94,45.64){\oval(2.67,2.67)[lt]}
\put(142.94,48.31){\oval(2.67,2.67)[rb]}
\put(120.28,22.97){\oval(2.67,2.67)[lt]}
\put(120.28,25.63){\oval(2.67,2.67)[rb]}
\put(117.61,20.30){\oval(2.67,2.67)[lt]}
\put(117.61,22.97){\oval(2.67,2.67)[rb]}
\put(114.94,17.63){\oval(2.67,2.67)[lt]}
\put(114.94,20.30){\oval(2.67,2.67)[rb]}
\put(112.28,14.97){\oval(2.67,2.67)[lt]}
\put(112.28,17.63){\oval(2.67,2.67)[rb]}
\put(109.61,12.30){\oval(2.67,2.67)[lt]}
\put(109.61,14.97){\oval(2.67,2.67)[rb]}
\put(109.61,58.97){\oval(2.67,2.67)[lb]}
\put(109.61,56.31){\oval(2.67,2.67)[rt]}
\put(112.27,56.30){\oval(2.67,2.67)[lb]}
\put(112.28,53.64){\oval(2.67,2.67)[rt]}
\put(114.94,53.64){\oval(2.67,2.67)[lb]}
\put(114.95,50.97){\oval(2.67,2.67)[rt]}
\put(117.61,50.97){\oval(2.67,2.67)[lb]}
\put(117.61,48.31){\oval(2.67,2.67)[rt]}
\put(120.27,48.30){\oval(2.67,2.67)[lb]}
\put(120.28,45.64){\oval(2.67,2.67)[rt]}
\put(142.94,25.64){\oval(2.67,2.67)[lb]}
\put(142.95,22.97){\oval(2.67,2.67)[rt]}
\put(145.61,22.97){\oval(2.67,2.67)[lb]}
\put(145.61,20.31){\oval(2.67,2.67)[rt]}
\put(148.27,20.30){\oval(2.67,2.67)[lb]}
\put(148.28,17.64){\oval(2.67,2.67)[rt]}
\put(150.94,17.64){\oval(2.67,2.67)[lb]}
\put(150.95,14.97){\oval(2.67,2.67)[rt]}
\put(153.61,14.97){\oval(2.67,2.67)[lb]}
\put(153.61,12.31){\oval(2.67,2.67)[rt]}
\put(147.26,51.86){\vector(1,1){1.11}}
\put(147.48,19.50){\vector(1,-1){1.11}}
\put(115.55,19.15){\vector(1,1){1.11}}
\put(115.59,52.28){\vector(1,-1){1.11}}
\put(86.94,56.31){\oval(2.67,2.67)[lt]}
\put(86.94,58.97){\oval(2.67,2.67)[rb]}
\put(84.27,53.64){\oval(2.67,2.67)[lt]}
\put(84.27,56.31){\oval(2.67,2.67)[rb]}
\put(81.60,50.97){\oval(2.67,2.67)[lt]}
\put(81.60,53.64){\oval(2.67,2.67)[rb]}
\put(78.94,48.31){\oval(2.67,2.67)[lt]}
\put(78.94,50.97){\oval(2.67,2.67)[rb]}
\put(76.27,45.64){\oval(2.67,2.67)[lt]}
\put(76.27,48.31){\oval(2.67,2.67)[rb]}
\put(53.61,22.97){\oval(2.67,2.67)[lt]}
\put(53.61,25.63){\oval(2.67,2.67)[rb]}
\put(50.94,20.30){\oval(2.67,2.67)[lt]}
\put(50.94,22.97){\oval(2.67,2.67)[rb]}
\put(48.27,17.63){\oval(2.67,2.67)[lt]}
\put(48.27,20.30){\oval(2.67,2.67)[rb]}
\put(45.61,14.97){\oval(2.67,2.67)[lt]}
\put(45.61,17.63){\oval(2.67,2.67)[rb]}
\put(42.94,12.30){\oval(2.67,2.67)[lt]}
\put(42.94,14.97){\oval(2.67,2.67)[rb]}
\put(42.94,58.97){\oval(2.67,2.67)[lb]}
\put(42.94,56.31){\oval(2.67,2.67)[rt]}
\put(45.60,56.30){\oval(2.67,2.67)[lb]}
\put(45.61,53.64){\oval(2.67,2.67)[rt]}
\put(48.27,53.64){\oval(2.67,2.67)[lb]}
\put(48.28,50.97){\oval(2.67,2.67)[rt]}
\put(50.94,50.97){\oval(2.67,2.67)[lb]}
\put(50.94,48.31){\oval(2.67,2.67)[rt]}
\put(53.60,48.30){\oval(2.67,2.67)[lb]}
\put(53.61,45.64){\oval(2.67,2.67)[rt]}
\put(76.27,25.64){\oval(2.67,2.67)[lb]}
\put(76.28,22.97){\oval(2.67,2.67)[rt]}
\put(78.94,22.97){\oval(2.67,2.67)[lb]}
\put(78.94,20.31){\oval(2.67,2.67)[rt]}
\put(81.60,20.30){\oval(2.67,2.67)[lb]}
\put(81.61,17.64){\oval(2.67,2.67)[rt]}
\put(84.27,17.64){\oval(2.67,2.67)[lb]}
\put(84.28,14.97){\oval(2.67,2.67)[rt]}
\put(86.94,14.97){\oval(2.67,2.67)[lb]}
\put(86.94,12.31){\oval(2.67,2.67)[rt]}
\put(80.59,51.86){\vector(1,1){1.11}}
\put(80.81,19.50){\vector(1,-1){1.11}}
\put(48.88,19.15){\vector(1,1){1.11}}
\put(48.92,52.28){\vector(1,-1){1.11}}
\end{picture}
\caption{ Four-quark condensate contribution to the imaginary
part of the non-forward $\gamma\gamma$-scattering amplitude.}
\end{figure}

\vspace{1cm}

\begin{minipage}[t]{7.8cm} {
\begin{center}
\vspace{5cm}
\hspace{3cm}\mbox{
\epsfysize=8.0cm
\epsffile[0 0 400 400]{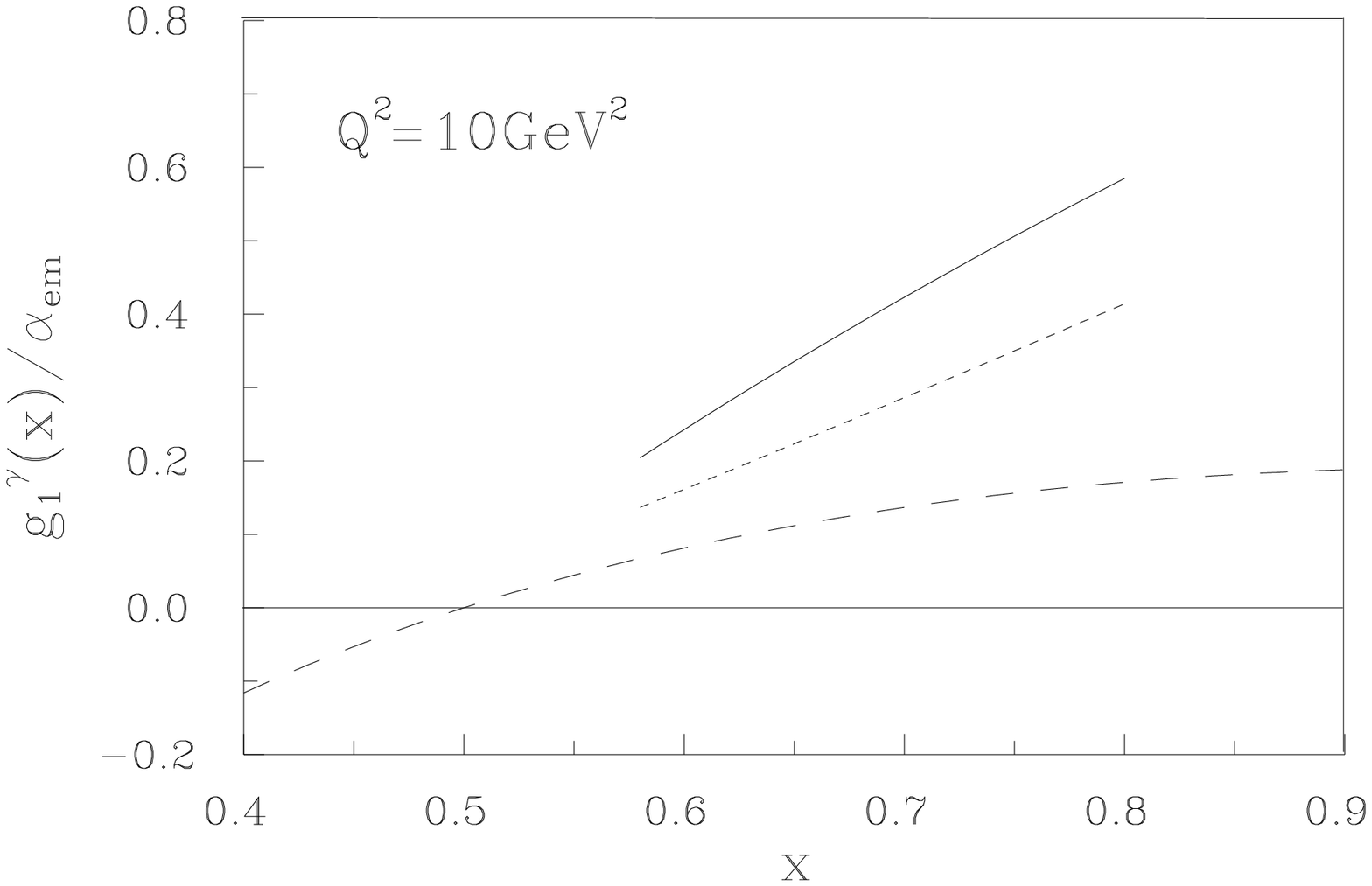}
}
\end{center}
}\end{minipage}

\vspace{0.5cm}

{\bf Fig.~5.}{ Spin dependent structure function of
the real photon at $Q^2=10GeV^2$. The solid curve corresponds
to the full structure function given by eq. (11), while long-
and short-dashed lines correspond to the continuum and hadronic
contributions to the latter.}

\end{document}